\documentclass[prb,showpacs,preprintnumbers,amsmath,amssymb,floatfix]{revtex4}
\usepackage[dvips]{graphicx}
\usepackage[latin1]{inputenc}
\begin{document}
\draft

\newcommand{\cu} {$^{63}$Cu }
\newcommand{\cuiso} {$^{63,65}$Cu }
\newcommand{\etal} {{\it et al.} }
\newcommand{\ie} {{\it i.e.} }
\newcommand{\aucr}{CeCu$_{5.9}$Au$_{0.1}$ }
\newcommand{\auaf}{CeCu$_{5.2}$Au$_{0.8}$ }
\newcommand{\aux}{CeCu$_{6-x}$Au$_{x}$ }
\newcommand{\ip}{${\cal A}^2$ }

\hyphenation{a-long}

\title{Mesoscopic phase separation in Na$_x$CoO$_2$ ($0.65\leq x\leq 0.75$)}

\author{P. Carretta, M. Mariani, C.B. Azzoni and M.C. Mozzati}
\affiliation{Department of Physics ``A.Volta" and Unit\`a INFM,
University of Pavia, Via Bassi 6, I-27100, Pavia (Italy)}
\author{I. Bradari\'c}
\affiliation{Laboratory for Theoretical and Condensed Matter
Physics, The "Vin\v{c}a" Institute of Nuclear Sciences, P.O. Box
522, 11001 Belgrade, Serbia}
\author{I. Savi\'c}
\affiliation{ Faculty of Physics, University of Belgrade,
Studentski trg 12-14, 11000 Belgrade, Serbia}
\author{A. Feher }
\affiliation{Faculty of Science, P. J. Safárik University, Park
Angelinum 9, 04154 Kosice, Slovakia }
\author{J. \v{S}ebek}
\affiliation{Institute of Physics, AS CR, Na Slovance 2, 18221
Prague 8, Czech Republic }
\widetext

\begin{abstract}

NMR, EPR and magnetization measurements in Na$_x$CoO$_2$ for
$0.65\leq x\leq 0.75$ are presented. While the EPR signal arises
from Co$^{4+}$ magnetic moments ordering at $T_c\simeq 26$ K,
$^{59}$Co NMR signal originates from cobalt nuclei in metallic
regions with no long range magnetic order and characterized by a
generalized susceptibility typical of strongly correlated metallic
systems. This phase separation in metallic and magnetic insulating
regions is argued to occur below $T^*(x)$ ($220 - 270$ K). Above
$T^*$ an anomalous decrease in the intensity of the EPR signal is
observed and associated with the delocalization of the electrons
which for $T<T^*$ were localized on Co$^{4+}$ $d_{z^2}$ orbitals.
It is pointed out that the in-plane antiferromagnetic coupling
$J\ll T^*$ cannot be the driving force for the phase separation.

\end{abstract}

 \maketitle

\narrowtext

\section{Introduction}

Na$_x$CoO$_2$ has been subject of an intense research activity in
the past owing to its possible technological applicabilities as a
battery electrode material \cite{battery}. More recently it has
attracted a renewed interest in view of its rich phase diagram and
for the several aspects it shares in common with the
superconducting cuprates \cite{Baskaran}. In particular, it has a
layered structure formed piling up CoO$_2$ layers containing
Co$^{4+}$ $S=1/2$ ions \cite{structure} and it becomes
superconducting when the chemical pressure is modified by
intercalating water molecules between Na and CoO$_2$ planes
\cite{superc}. On the other hand, while in the cuprates Cu$^{2+}$
ions form a square lattice, in Na$_x$CoO$_2$ cobalt ions form a
triangular lattice which induces geometrical frustration of the
antiferromagnetic interactions \cite{Pelobianco}. Moreover,
Na$_x$CoO$_2$, for $x$ around 0.7 shows an anomalous phase
transition at $T_c\simeq 26$ K, evidenced by a small peak in the
specific heat. The low-temperature (T) ground state yields an
extremely small macroscopic magnetization \cite{Moto} and it was
argued \cite{muSR}, on the basis of $\mu$SR measurements, that it
might correspond to a spin-density wave (SDW). Furthermore,
slightly below room temperature, on the basis of NMR measurements
alone, Gavilano et al. Ref.\onlinecite{Gavilano} suggested that
charge ordering occurs. In this T range anomalies in $^{59}$Co NMR
spectra \cite{NMRRay} and a crossover from insulating to
metallic-like behavior in the c-axes resistivity \cite{resist} was
observed for $x\simeq 0.5$. Finally, a recent analysis of low-T
thermal and transport properties suggested that for $x$ around 0.7
the electronic properties have to be described by a two-band model
\cite{Batlogg}.

In the following, based on the analysis of NMR and EPR data, it
will be shown that metallic and magnetic insulating domains
coexist in Na$_x$CoO$_2$ for $0.65\leq x\leq 0.75$. In particular,
while EPR is sensitive to the magnetic domains with electrons
localized on Co$^{4+}$ $d_{z^2}$ orbitals, NMR spectra and
relaxation measurements allow to investigate the spin dynamics in
the metallic regions. The phase separation is argued to occur
below $T^*(x)$ ($220- 270$ K), the same temperature at which
Gavilano et al. Ref.\onlinecite{Gavilano} suggested the occurrence
of charge ordering. Above $T^*$ the electrons which were localized
on Co$^{4+}$ $d_{z^2}$ orbitals can delocalize through an
activated process.

In the next section the experimental results and technical aspects
involved in the sample preparation, magnetization, EPR and NMR
measurements will be given. The analysis of the EPR and NMR
spectra providing evidence for a phase separation will be
presented in Sect.III, together with the analysis of the
generalized spin susceptibility within the metallic phase. The
concluding remarks will be given in Sect. IV.

\section{Experimental results}

\subsection{Sample Preparation}

Na$_x$CoO$_2$ ($x=0.65, 0.70$ and $0.75$) samples were prepared
following the "rapid heat-up" method \cite{Bra1,Bra2}. A
stoichiometric mixture of 99.99\% purity Co$_3$O$_4$ and
Na$_2$CO$_3$ was thoroughly ground, and placed in a furnace and
preheated at 750 C for 12 hours. The obtained samples were
reground and annealed for 15 hours at 850 C in air, followed by
slow cooling to room temperature. X-ray powder diffraction
measurements confirmed all samples to be single phase of hexagonal
$\gamma$-Na$_x$CoO$_2$ \cite{Bra3,Bra4}. The lattice parameter $a$
was estimated around 2.827 \AA \, for all samples, while $c=
10.939(3), 10.907(4)$ and $10.892(4)$ \AA \, for $x=0.65, 0.70$
and $0.75$, respectively, in good agreement with the values
reported in literature (see Fig.\ref{XRD}).

\subsection{Magnetization}

Magnetization  measurements were performed using a Quantum Design
XPMS-XL7 SQUID magnetometer. The field cooled magnetization (M)
for $T\geq T_c\simeq 26$ K was observed to increase linearly with
the magnetic field intensity (H). The temperature dependence of
the susceptibility $\chi =M/H$, after correcting for the core
electrons diamagnetism, for $x=0.65, 0.7$ and $0.75$, is reported
in Fig.\ref{FigchivsT}. One notices that $\chi$ increases on
cooling and shows a small kink at $T_c\simeq 26$ K in all samples.
The magnitude of $\chi$ was observed to increase sizeably upon
decreasing $x$. The susceptibility measurements were repeated on
the $x=0.75$ sample after $4$ months and no ageing effect was
noticed, except for a small increase in the low-temperature
susceptibility due to possible contamination from paramagnetic
impurities.

\subsection{EPR measurements}

EPR spectra were recorded with an X-band spectrometer equipped
with a standard microwave cavity and a variable temperature
device. The room temperature derivative of the EPR powder spectra
for $x=0.65, 0.7$ and $0.75$ is shown in Fig.\ref{spEPR}. The
spectra are broad and slightly asymmetric, with a $g\simeq 2$,
typical of Co$^{4+}$ ions with a distorted octahedral coordination
and a low-spin $S=1/2$ configuration. One notices a remarkable
decrease in the intensity of the EPR signal with increasing Na
content. The intensity of the EPR signal was calibrated with
respect to the one of a reference paramagnetic salt and it was
found that even for $x=0.65$ only about $8$\% of all Co sites
contribute to the EPR signal. This reduced intensity at room
temperature should not be associated with a poor penetration of
the microwave inside the grains. In fact, at room temperature the
estimated skin depth is $18$ $\mu$m, close to the average grain
size. Only at low temperature the shortening of the skin depth can
lead to a poor irradiation. The low EPR signal originates from a
reduced fraction of Co$^{4+}$ sites. The intensity of the EPR
signal (see Fig.\ref{spEPRvsT}), which in principle is
proportional to the contribution  to the static uniform
susceptibility of the irradiated Co$^{4+}$ ions, shows an
anomalous decrease above $T^*$ ($\simeq 240$K for $x=0.7$ and
$\simeq 270$ K for $x=0.65$) and vanishes around $500$ K. At low
temperature the signal intensity passes through a maximum around
$125$ K, then decreases and vanishes abruptly at $T_c$, indicating
that at this temperature a transition to a phase with long range
order among Co$^{4+}$ moments occurs. The temperature dependence
of the EPR linewidth shows a minimum around 115 K characteristic
of two-dimensional (2D) antiferromagnets \cite{Richards,MnCl2}
(see the inset to Fig.\ref{spEPR}).

\subsection{NMR Spectra and relaxation}

NMR spectra and spin-lattice relaxation measurements were carried
out using standard radio-frequency (RF) pulse sequences. $^{23}$Na
echo signal was maximized with a $\pi/2 -\tau - \pi$ pulse
sequence while $^{59}$Co echo signal was maximized with a $\pi/2
-\tau -\pi/2$ sequence. The $^{23}$Na (see Fig.\ref{spna}) and
$^{59}$Co NMR powder spectra were obtained from the Fourier
transform of half of the echo signal and from the envelope of the
echo magnitude, respectively.

$^{23}$Na NMR spectra of the central line, associated with the
$1/2 - -1/2$ transition, were observed to progressively shift to
higher frequency and to broaden on cooling (see
Fig.\ref{shiftNa}). However, the shape of the low-T spectra was
not reproducible. Even after a temperature cycle in helium
atmosphere, where the sample was warmed from $35$ K up to $280$ K
and after 50 minutes cooled back to $35$ K, a change in the
spectrum was noticed (see Fig.\ref{spna}). This evidences that the
modifications are not due to ageing effects but rather reflect
intrinsic differences in the microscopic environment around the
nuclei. In some cases a nearly symmetric line shape was observed
whereas in other measurements two distinct peaks were visible
\cite{NMRStall} (see Fig.\ref{spna}). The two peaks of $^{23}$Na
central line should not be ascribed to the two singularities
expected for a quadrupolar perturbed NMR powder spectrum
\cite{Abragam} since, as will be shown hereafter, the temperature
dependence of the spin-lattice relaxation rate measured at the two
peaks is different. Also $^{59}$Co NMR spectra show analogous
modifications after the sample has undergone a temperature cycle
between 35 K and  room temperature.

It must be mentioned that neither $^{59}$Co nor $^{23}$Na spectra
show any sizeable change when the sample is cooled below $T_c$. As
can be seen in Fig.\ref{spCoTc} the $^{59}$Co NMR spectrum of the
central line is practically identical above and below $T_c$.{\it
This is a clear indication that the $^{59}$Co nuclei giving rise
to the NMR signal, corresponding to the majority of cobalt nuclei,
are in regions with no long range magnetic order}. It must be
mentioned that the observation of $^{59}$Co nuclei belonging to
Co$^{4+}$ ions, the ones yielding the EPR signal, is prevented by
the extremely fast nuclear relaxation. Hence, EPR and NMR in
Na$_x$CoO$_2$ are complementary. The first probes the local
susceptibility of Co$^{4+}$ rich regions which show a long range
magnetic order below $26$ K, the second allows to investigate the
static and dynamic magnetic properties of the metallic regions
with non-magnetic cobalt ions.

The paramagnetic shift $\Delta K$ of $^{23}$Na NMR central line
shows the same temperature dependence of $\chi$ measured with the
SQUID magnetometer. In fact, one can write
\begin{equation}
\label{eq:A}
          \Delta K = {A \chi({\bf q}=0,\omega =0)\over g\mu_BN_A} + \delta
\end{equation}
with $A$ the hyperfine coupling, $\chi(0,0)$ the static uniform
molar spin susceptibility and $\delta$ the chemical shift. Hence,
the plot of $\Delta$K vs. $\chi$ directly yields the hyperfine
coupling (see Fig.\ref{DKChi}). For $x=0.65$ $A\simeq 16$ kOe,
while for $x=0.75$ a slightly smaller value $A\simeq 14$ kOe was
found.

Nuclear spin-lattice relaxation rate $1/T_1$ was measured by using
a saturation recovery pulse sequence. The recovery of the nuclear
magnetization after saturation of the central line deviates a bit
from the expected behavior, namely $[1-m(t)/m(t\rightarrow\infty
)]= 0.1 exp(-t/T_1) + 0.9 exp(-6t/T_1)$ for $^{23}$Na and
$[1-m(t)/m(t\rightarrow\infty )]= 0.012 exp(-t/T_1) + 0.068
exp(-6t/T_1) + 0.206 exp(-15t/T_1) + 0.714 exp(-28t/T_1)$ for
$^{59}$Co, with $m(t)$ the nuclear magnetization recovered at time
$t$ after the saturation sequence. This discrepancy is due to the
fact that the different parts of $^{59}$Co and $^{23}$Na NMR
spectra have relaxation rates which differ both in magnitude and
also in regards of the temperature dependence. This is evident,
for instance, when the recovery of the intensity of the two
shoulders of $^{23}$Na NMR spectrum in Fig.\ref{spna} is recorded
(see Fig.\ref{recNa}). One notices a fast relaxing component,
associated with the less intense shoulder, and a slow relaxing
component due to the most intense peak. The T-dependence of the
relaxation rate of the fast relaxing $^{23}$Na nuclei shows a
small kink at $T_c$ (see the inset to Fig.\ref{T1Na}) and should
be associated with those nuclei close to Co$^{4+}$ ions. On the
other hand, the T-dependence of $1/T_1$ due to the slow relaxing
nuclei shows a different trend and no peak at $T_c$. These nuclei
are the ones belonging to the metallic regions. Unfortunately, it
is possible to discern these two components only at low
temperature and only when $^{23}$Na NMR spectra shows well defined
shoulders. When this separation is not possible, since most of
$^{23}$Na nuclei are characterized by a slow relaxation rate, the
measured $1/T_1$ probes essentially the spin excitations of the
metallic regions of the sample (see Fig.\ref{T1Na}). $1/T_1$ is
observed to increase smoothly with temperature up to about 100 K
where a more pronounced increase is observed, followed by an
abrupt increase around $T^*\simeq 220$ K for $x=0.75$ and around
$T^*\simeq 270$ K for $x=0.65$, similarly to what was observed by
Gavilano et al. Ref.\onlinecite{Gavilano}. The temperature
dependence of $^{59}$Co $1/T_1$ shows a similar behavior and no
anomaly at $T_c$ (see Fig.\ref{t1alpa}).

\section{Analysis and Discussion}

First we will concentrate on the experimental evidences for a
phase separation in metallic and in insulating magnetic domains
and then we will discuss in more detail the temperature dependence
of the static uniform susceptibility  probed by NMR shift and
magnetization measurements and of the low-frequency dynamics
investigated by means of $1/T_1$ measurements.

\subsection{Phase separation}

As already mentioned in the previous section the EPR signal is
characteristic of Co$^{4+}$ ions in a low-spin state, with the
lowest $t_{2g}$ levels fully occupied and the highest $a_1^T$
(corresponding to $d_{z^2}$) orbital half-filled
\cite{Good,Baskaran}. The degeneracy among the lowest
twofold-degenerate $e^T$ and the $a_1^T$ levels is relieved by the
trigonal distortion around cobalt. Crystal field calculations show
that these levels are well separated and hence $g\simeq 2$ and
that a reduced spin-orbit coupling can account for the small
anisotropy of the $g$ tensor yielding the slightly asymmetric EPR
line (see Fig.\ref{spEPR}). For $T>T^*$ the electrons localized on
the $a_1^T$ orbital can be promoted to the conduction band formed
by the hybridization of Co$^{4+}$ $e^T$ and O $2p$ orbitals,
through an activated process. This leads to an activated transport
along the $c$-axes \cite{resist} and to a polaronic-like motion of
the electrons, as pointed out by Rivadulla et al.
Ref.\onlinecite{Good}. When the electrons are promoted to the
conduction band the number of Co$^{4+}$ sites is progressively
reduced, the intensity of the EPR signal diminshes with
temperature (see Fig.\ref{spEPR}) and, finally, for $T
> 500$ K the majority of the electrons are itinerant. Hence, one
can derive the T-dependence of Co$^{4+}$ sites density directly
from the EPR data.

In order to estimate the number of Co$^{4+}$ sites one has first
to take into account that the EPR signal intensity depends on the
T-dependent static uniform susceptibility and on the microwave
screening, which is relevant at low-T. The T-dependence of the
spin susceptibility was assumed to be the one of a 2D triangular
antiferromagnet \cite{ChiTALAF}, with a Curie-Weiss temperature
$\Theta\simeq 135$ K. This value is suggested by the minimum in
the T-dependence of the EPR linewidth, which usually occurs at a
temperature slightly below $\Theta$ \cite{Richards,MnCl2}. The
maximum in the T-dependence of the EPR intensity around $125$ K is
not the one characteristic of the susceptibility of a 2D
triangular antiferromagnet, which should occur around $0.35
\Theta\simeq 47$ K (see the dotted line in Fig.\ref{spEPRvsT}).
Also the decrease of the EPR intensity below $100$ K is too fast
to originate from the T dependence of the susceptibility of a 2D
triangular antiferromagnet. In fact, the pronounced reduction of
the low-T EPR intensity results from the decrease of the
skin-depth $d$ with temperature, which is proportional to the
square-root of the electrical resistivity. In order to take into
account this effect we have assumed for simplicity spherical
grains, with an average radius of $30$ $\mu$m, and that a grain is
fully irradiated over a distance equal to $d$ ($18$ $\mu$m at room
temperature). Taking into account the T-dependence of
Na$_x$CoO$_2$ resistivity reported in literature we have derived
the curve shown in Fig.\ref{spEPRvsT} for the expected integrated
EPR intensity. The experimental data are observed to follow rather
well the expected behavior below $T^*$, pointing out that for
$T\leq T^*$ the density of Co$^{4+}$ sites is T-independent. The
reduction of the EPR intensity above $T^*$, with respect to the
estimated one allows to derive the temperature evolution of the
fraction of electrons localized on Co$^{4+}$ $a_1^T$ orbitals (see
Fig.\ref{fracCo}). If $\Delta$ is the energy difference between
the $a_1^T$ level and the conduction band, the statistical
population of Co$^{4+}$ ions should be given roughly by
$n_{Co^{4+}}(T)/n_{Co^{4+}}(T\ll T^*)=1/(1+
N_{eff}exp(-\Delta/T))$, with $N_{eff}$ an effective density of
states for the itinerant electrons. Although the data in
Fig.\ref{fracCo} can be fitted rather well for a value of
$\Delta\simeq 0.3$ eV a more quantitative estimate would require
the knowledge of the T-dependence of $\Delta$ and therefore this
value should be taken just as an order of magnitude.

The observation of an EPR signal below T$^*$ suggests that,
although the electrical resistivity has a metallic behavior
\cite{Batlogg}, there is still a sizeable fraction of electrons
localized on Co$^{4+}$ $a_1^T$ orbitals at low T and, hence, $T^*$
cannot signal a conventional metal to insulator transition
\cite{Good}. The absolute value of the fraction of Co$^{4+}$ sites
for $T\ll T^*$, estimated from the EPR measurements, is around 8\%
for $x=0.65$ and progressively decreases upon increasing the Na
content. This trend is exactly the one of the magnetic entropy
estimated from specific heat measurements carried out on samples
of the same batch \cite{Feher} (see Fig.\ref{entropy}). In
particular, the $x$ dependence of the intensity of the specific
heat peak at $T_c$ scales with the EPR intensity upon increasing
$x$ (see the inset to Fig.\ref{entropy}). Moreover, the magnitude
of the magnetic entropy \cite{Feher} also indicates that for
$x=0.65$ a fraction around 8\% of $S=1/2$ ions contribute to the
total entropy. This is a neat confirmation that {\it the magnetic
transition yielding the specific heat peak at $26$ K is due to the
ordering of localized Co$^{4+}$ magnetic moments}. Also $\mu$SR
measurements suggest that only a small fraction of the sample
becomes magnetic. In fact, Sugiyama et al. Ref.\onlinecite{muSR}
observed that for $x=0.75$ no more than 20 \% of the muons
injected in the sample go into (or close to) magnetically ordered
regions for $T\leq T_c$. This is not by itself an indication of a
SDW phase since in that case, although one might expect a reduced
value of the sample magnetization, all the sample should show a
long-range order.

On the other hand, $^{59}$Co NMR relaxation measurements and
spectra, due to cobalt nuclei in metallic regions, show no sign of
magnetic ordering. {\it These observations clearly point towards a
phase separation between antiferromagnetic insulating and metallic
regions for $T<T^*$}. Above $T^*$ no phase separation occurs and
the electrons which for $T<T^*$ were localized  on the $a_1^T$
orbital can move with a polaronic-like motion through all the
sample \cite{Good}.

The difference in the NMR spectra after temperature cycling above
$T^*$ can be due to a different topological arrangement of the
insulating and metallic domains. In the case of $^{23}$Na NMR
spectra, the fast relaxing nuclei yielding the high frequency
shoulder (see Fig.\ref{spna})) are the ones closer to Co$^{4+}$
ions, while the most intense peak should be associated with
$^{23}$Na nuclei well inside the metallic regions. Since no effect
of thermal cycles was noticed on the intensity of the EPR signal
the ratio of the insulating over metallic volume should not change
after the temperature cycles. Then, a different ratio in the
intensity of the two components of $^{23}$Na NMR spectra should
originate from a modification in the surface/volume ratio of the
insulating regions. An increase in the intensity of the high
frequency shoulder should be caused by an average decrease in the
size of the insulating domains and vice-versa. Now, why the
topology of the metallic and insulating domains should be affected
by the temperature cycles? One possibility is that at high
temperatures, due to their relatively high mobility, Na$^+$ ions
modify their arrangement and lead to a modification in the Coulomb
potential. In particular, one should expect that the metallic
domains are attracted by Na$^+$ ionic potential, while the
Co$^{4+}$ rich insulating domains are repulsed.

A lower boundary for the size $L$ of the magnetic domains can be
estimated by taking into account that $T_c$ is $x$-independent
and, therefore, is not affected by finite size effects. Then at
$T_c$ the in-plane magnetic correlation length
$\xi_{Co^{4+}}(T_c)\ll L$. By taking for $\xi_{Co^{4+}}(T)$ the
T-dependence expected for a 2D triangular antiferromagnet and the
in-plane exchange coupling $J= 2\Theta/3\simeq 90$ K, one finds
\cite{ChiTALAF} that at $T_c$ the correlation length is less than
$3$ lattice steps. $\xi_{Co^{4+}}(T_c)$ can be estimated also by
taking a mean field expression for $T_c\simeq
J_{\perp}\xi^2_{Co^{4+}}(T_c)$, with $J_{\perp}\simeq 10^{-2}
J\simeq 0.9 K$ \cite{resist} the interplanar exchange coupling
among Co$^{4+}$ magnetic moments. One finds
$\xi_{Co^{4+}}(T_c)\simeq 5.4$ lattice steps. Hence, it must be
concluded that $L\gg 6$ lattice steps. Although an upper estimate
for $L$ cannot be made it must be remarked that one can conclude
that the phase separation is not macroscopic, i.e. due to chemical
inhomogeneities, but mesoscopic. First of all, it would be rather
singular that all data reported in literature on Na$_x$CoO$_2$
samples prepared in different ways indicate the same $T_c$ and a
similar magnitude of specific heat peaks \cite{Moto,Batlogg}.
Second, if the insulating magnetic domains where macroscopic the
EPR signal intensity should not suffer from microwave irradiation
problems. Third, the intensity of $^{23}$Na and $^{59}$Co NMR
spectra should not be affected by thermal cycles if the separation
was macroscopic.

Finally, it is important to observe that the phase separation is
already present at $T\gg J$. This clearly indicates that the
antiferromagnetic coupling among Co$^{4+}$ ions cannot be the
driving force for this phase separation and that alternative
explanations should be envisaged \cite{Grilli}.

\subsection{Spin dynamics in the metallic domains}

The magnetization and NMR data allow to probe the static and
dynamic magnetic properties of the metallic regions which do not
undergo a phase transition. The total static uniform
susceptibility measured with the SQUID is the sum of two terms
$\chi= \chi_{Co^{4+}} + \chi_{NFL}$, the first due to the magnetic
domains and the second due to the metallic ones. One notices (see
Fig.\ref{FigchivsT}) that $\chi_{Co^{4+}}$ is rather small and can
account just for the small shift in the high T absolute value of
$\chi(x)$, but not for the sizeable increase on cooling. In fact,
also in the $x=0.75$ sample, where $\chi_{Co^{4+}}$ is negligible,
an increase of $\chi$ upon cooling is observed. This increase
indicates that strong correlations among the electrons are present
and that the metallic regions cannot be treated as a weakly
correlated Fermi liquid. Also specific heat $C$ measurements show
a low temperature upturn in $C/T$ \,\, \cite{Batlogg}
characteristic of 2D electron systems with antiferromagnetic
correlations \cite{vonL}. Following Moriya et al.
Ref.\onlinecite{Moriya} one would expect a high-T Curie-Weiss
behavior both for the uniform static susceptibility $\chi(0,0)$
and in $1/T_1T$. However, if one plots $1/\chi(0,0)$ vs. T, with
$\chi(0,0)$ the one measured with the SQUID one observes a
non-linear behavior (see Fig.\ref{chialpa}). We remark that here
there is no sense in subtracting a T-independent Pauli-like
contribution since the electrons are strongly correlated. Also if
one plots the inverse of the spin contribution to $^{23}$Na NMR
shift $\Delta K= A\chi(0,0)/g\mu_BN_A$ one observes a non-linear
temperature dependence. Therefore, one has to resort to a form of
the generalized spin susceptibility which can describe the
non-Fermi-liquid (NFL) behavior. Recently, based on inelastic
neutron scattering results, combined with heuristic arguments a
response function of two-dimensional (2D) character has been used
to describe the NFL behavior  in the proximity of a quantum
critical point \cite{uno,due}:
\begin{equation}
\label{eq:1}
          \chi_{NFL}^{-1}\,({\bf q},\omega,T)=
          k_B
          \left[\frac{(T-i\omega/a)^{\alpha}}{c}
          +f^{\alpha} ({\bf q},T)
          \right]
\end{equation}
with an anomalous exponent $\alpha\neq 1$, $(\omega/T)$ scaling
and renormalized Curie-Weiss constant $c$. $f^{\alpha} ({\bf
q},T)=(aT)^{\alpha}(1+ q^2\xi^2)$ (with ${\bf q}$ starting from
the critical wavevector ${\bf Q}$), $(aT)^{\alpha}=\Gamma_{\bf Q}$
the critical frequency and $\xi(T)=\sqrt{\Theta' /T \vert {\bf
Q}\vert^2}$ the correlation length. In particular a value
$\alpha\simeq 0.7$ was observed to justify the behavior of
CeCu$_{5.9}$Au$_{0.1}$ generalized susceptibility \cite{uno,tre}.
In the light of these findings one can resort to the above form of
$\chi_{NFL}({\bf q},\omega)$ to analyze magnetization and $1/T_1$
results. From Eq.\ref{eq:1} one finds that
\begin{equation}
\label{eq:2}
          \chi_{NFL}(0,0,T)= {c\over T^{\alpha} + \Theta '^{\alpha}} .
\end{equation}
Then, for a certain value of $\alpha$ by plotting $1/\chi(0,0)$
vs. $T^{\alpha}$ one should observe a linear behavior. Remarkably
this was observed to occur for $\alpha=0.7$ (see
Fig.\ref{chialpa}), as in CeCu$_{5.9}$Au$_{0.1}$ \cite{vonL}. The
values derived for $\Theta'$ for the different Na contents are
close to $75$ K.

Now we turn to the discussion of the spin-lattice relaxation rate.
Below $T^*$ one observes a drastic decrease in the amplitude of
$1/T_1$. In particular, $1/T_1T$ decreases down to about $100$ K
and then starts to increase again, as expected for a NFL. $1/T_1$
was found to be field independent up to T$^*$\,\, \cite{Gavilano},
therefore, the drastic decrease below $T^*$ does not indicate a
slowing down of the fluctuations to frequencies below nuclear
Larmor frequency $\omega_L$ but either a decrease in the amplitude
of the spin fluctuations or a drastic increase in the frequency of
the fluctuations. This can be expected in correspondence of a
crossover from a high temperature rather slow dynamics probed by
$1/T_1$, associated with polaronic-like motions of the electrons,
to a low-T fast dynamics of the itinerant electrons. Below $100$ K
the behavior is the one expected for a NFL and one can verify if,
by using the form of $\chi_{NFL}({\bf q},\omega)$ reported above
and $\alpha=0.7$, the T-dependence of $1/T_1$ is reproduced. From
Eq.\ref{eq:1}, by recalling that \cite{Moro}
\begin{equation}
\label{eq:3} {1\over T_1}=
   \frac{{\gamma}^2}
   {2N}\,
   k_B\,T\,{\cal A}^2 \sum_{\bf q}
   \left[ \frac{\chi_{NFL} '' ({\bf q},\omega_L)}
   {\omega_L} \right]
\end{equation}
with $\gamma$ the nuclear gyromagnetic ratio, under the condition
that all the excitation frequencies $\Gamma_{\bf q}$ remain larger
than $\omega_L$ and in case of an external field $H<<
k_BT/g\mu_B$, one derives
\begin{equation}
\label{eq:4} {1\over T_1}=
   \frac{{\gamma}^2}
   {2N}\,
   k_B\,T\,{\cal A}^2
   \sum_{2D}^{BZ}
   \frac{c}{a^{\alpha}}\,\,
   \frac{\left[1+(q\,\xi)^{2\alpha}
   \right]^{-2}}{k_B\,T^{\alpha+1}} .
\end{equation}
All constants involved in Eq.\ref{eq:4}, except $\vert{\bf
Q}\vert$ (entering $\xi(T)$ expression) and $a$, have been
previously determined through the analysis of the NMR shift and of
the static uniform spin susceptibility. One observes (see
Fig.\ref{t1alpa}) that the T-dependence of $1/T_1$ can be
reproduced reasonably well for $\alpha=0.7$ by taking a value of
$a$ such that $\Gamma_{\bf Q}(T=10 K)\simeq 2.2\times 10^{12}$
s$^{-1}$ and $\xi(T=10 K)\simeq 9$ lattice steps, for $x=0.75$,
and $\Gamma_{\bf Q}(T=10 K)\simeq 1.3\times 10^{12}$ s$^{-1}$ and
for $\xi(T=10 K)\simeq 11$ lattice steps, for $x=0.65$.

\section{Conclusions}

It was shown through a series of EPR, NMR and magnetization
measurements that in Na$_x$CoO$_2$ for $0.65\leq x\leq 0.75$ phase
separation in metallic and in magnetic domains is present. The
phase separation develops below $T^*\gg J$ and therefore, it is
unlikely that the antiferromagnetic exchange is the driving
mechanism for the phase separation. Above T$^*$ the activated
delocalization of the electrons yields a marked reduction in the
EPR amplitude. The spin excitations in the metallic regions are
suitably described by an heuristic form of the susceptibility
successfully used to describe the properties of strongly
correlated metals. All the above observations are relevant in
regards of the phase separation observed in the cuprates since
they share many aspects in common with Na$_x$CoO$_2$. However, a
deep comparison of the two systems goes beyond the aim of this
work. Here we only mention that while the strong antiferromagnetic
coupling in the cuprates can be relevant for the phase separation
or for stripes formation \cite{Tranqu}, in Na$_x$CoO$_2$ it is
not. On the other hand, it is interesting to observe that in both
systems phase separation is stabilized for those doping levels $x$
yielding a charge distribution which is commensurate with the
lattice, namely $x\simeq 2/3$ for the 2D triangular lattice of
Na$_x$CoO$_2$ and $x=1/8$ for the 2D square lattice of the
cuprates \cite{Tranqu}.

\section*{Acknowledgement}

The research activity in Pavia was supported by the project
MIUR-FIRB {\it Microsistemi basati su materiali magnetici
innovativi strutturati su scalla nanoscopica}. I.Bradari\'c and
I.Savi\'c were supported by Serbian Ministry of Science,
Technology, and Development, Grant No. 1899. A. Feher and J. Sebek
acknowledge the support by Science and Technology Assistance
Agency under contract  Nos. APVT- 20-009902 and GAAC- 202/03/0550.


\begin{figure}
 \caption{The
$x$-dependence of the $c$-axes length is reported for the
Na$_x$CoO$_2$ powder samples investigated in this work and for
other samples reported in the literature.} \label{XRD}
\end{figure}

\begin{figure}
\caption{Temperature dependence of the susceptibility $\chi =M/H$,
for $H= 40$ Gauss, in Na$_x$CoO$_2$ powder samples ($x=0.65, 0.70$
and $0.75$). The dotted line shows the contribution to the
susceptibility from Co$^{4+}$ magnetic moments for $x=0.65$, as
derived from EPR measurements (see the dotted line in
Fig.\ref{spEPRvsT}). In the inset the field dependence of the
magnetization at $280$ K is shown for the $x=0.65$ sample. }
\label{FigchivsT}
\end{figure}

\begin{figure}
 \caption{Derivative of
the EPR signal in Na$_x$CoO$_2$ for $x=0.65, 0.70$ and $0.75$, at
$T=293$ K. The intensity was normalized to the same reference
value for all Na contents. In the inset the T-dependence of the
linewidth of the EPR absorption signal for $x=0.65$ is reported
(the line is a guide to the eye).} \label{spEPR}
\end{figure}

\begin{figure}
\caption{The T-dependence of the integrated EPR signal is reported
for the $x=0.65$ and $x=0.7$ Na$_x$CoO$_2$ powder samples. The
solid lines show the behavior expected on the basis of the
T-dependence of the static uniform susceptibility and of the
microwave skin-depth (see text). The dotted line shows the
behavior of the static uniform susceptibility which should be
followed if no shielding of the microwaves was present.}
\label{spEPRvsT}
\end{figure}

\begin{figure}
 \caption{$^{23}$Na
central line NMR powder spectrum in Na$_{0.75}$CoO$_2$, for $H= 6$
Tesla. {\it Top} The spectrum at $35$ K is reported before (dotted
line) and after (solid line) having warmed the sample at 280 K in
helium atmosphere for 50 minutes. {\it Bottom} The spectrum at
$16$ K is reported for two different runs performed few weeks
apart.} \label{spna}
\end{figure}

\begin{figure}
 \caption{T-dependence
of $^{23}$Na central line NMR shift in Na$_x$CoO$_2$ for $x=0.65$
and $x=0.75$. In the inset the T-dependence of the full width at
half intensity of $^{23}$Na central line is reported. These data
were taken when $^{23}$Na central line was nearly symmetric with
no shoulders.} \label{shiftNa}
\end{figure}

\begin{figure}
 \caption{$^{59}$Co
(left) and $^{23}$Na (right) central line NMR powder spectra in
Na$_{0.75}$CoO$_2$, for $H= 16$ kGauss, for T above and below
$T_c$. The intensity of ${23}$Na signal is reduced to about half
of its actual value since the length of the $\pi/2$ RF pulse,
calibrated on $^{59}$Co signal, was kept constant while sweeping
the frequency.} \label{spCoTc}
\end{figure}

\begin{figure}
 \caption{The
paramagnetic shift of $^{23}$Na central line is plotted as a
function of the macroscopic susceptibility, measured with a SQUID
magnetometer, for the $x=0.75$ Na$_x$CoO$_2$ sample. }
\label{DKChi}
\end{figure}

\begin{figure}
 \caption{The recovery
of $^{23}$Na central line Fourier transform intensity after a
saturation RF pulse ($y(t)= 1-m(t)/m(\infty)$) . The circles show
the recovery of the low frequency peak, whereas the squares show
the recovery of the high frequency shoulder (see Fig.\ref{spna}.}
\label{recNa}
\end{figure}

\begin{figure}
 \caption{T-dependence
of $^{23}$Na $1/T_1$ in Na$_{0.75}$CoO$_2$ for $H= 6$ Tesla,
derived from the recovery of the echo amplitude after a saturating
RF pulse. In the inset these $1/T_1$ values are compared to the
ones derived from the recovery of the Fourier transform
high-frequency peak (closed circles) or low-frequency shoulder
(open circles) (see Fig.\ref{spna} and Fig.\ref{recNa}). }
\label{T1Na}
\end{figure}

\begin{figure}
 \caption{T-dependence
of the fraction of Co$^{4+}$ ions in Na$_{0.65}$CoO$_2$,
normalized to its value for $T\ll T^*$. Above $T^*$ a marked
decrease is evident. The solid line shows the best fit for a gap
between localized and itinerant states $\Delta\simeq 0.3$ eV (see
text). In the inset the $x$-dependence of $T^*$ is reported with a
schematic view of the portion of the $x$-T phase diagram where
phase separation (PS) is observed.} \label{fracCo}
\end{figure}

\begin{figure}
 \caption{$x$-dependence
of the fraction of Co$^{4+}$ sites in Na$_x$Co$_2$, normalized to
its absolute value for $x=0.65$ ($n_{Co^{4+}}(0.65\simeq 0.08$).
The open diamonds show the values estimated from the integrated
EPR intensity while the closed circles show the values derived
from the estimated entropy due to localized $S=1/2$ spins
\cite{Feher}. In the inset the T-dependence of the ratio $C/T$,
with $C$ the specific heat, is reported. The decrease in the
specific heat at $T_c\simeq 26$K with increasing $x$ is evident.}
\label{entropy}
\end{figure}

\begin{figure}
\caption{The inverse of the spin susceptibility measured with a
SQUID magnetometer is reported against T (top) and $T^{0.7}$
bottom for $x=0.75$. The non-linearity in the former plot and the
linearity in the latter one are evident.} \label{chialpa}
\end{figure}

\begin{figure}
\caption{T-dependence of $^{59}$Co (top) and $^{23}$Na NMR
$1/T_1T$ in Na$_{0.75}$CoO$_2$, for $H= 6$ Tesla, compared to the
theoretical behavior (dotted line) expected for an exponent
$\alpha= 0.7$ (see text). The theoretical curve for $^{59}$Co was
derived using a hyperfine coupling constant $A=63$ kOe.}
\label{t1alpa}
\end{figure}

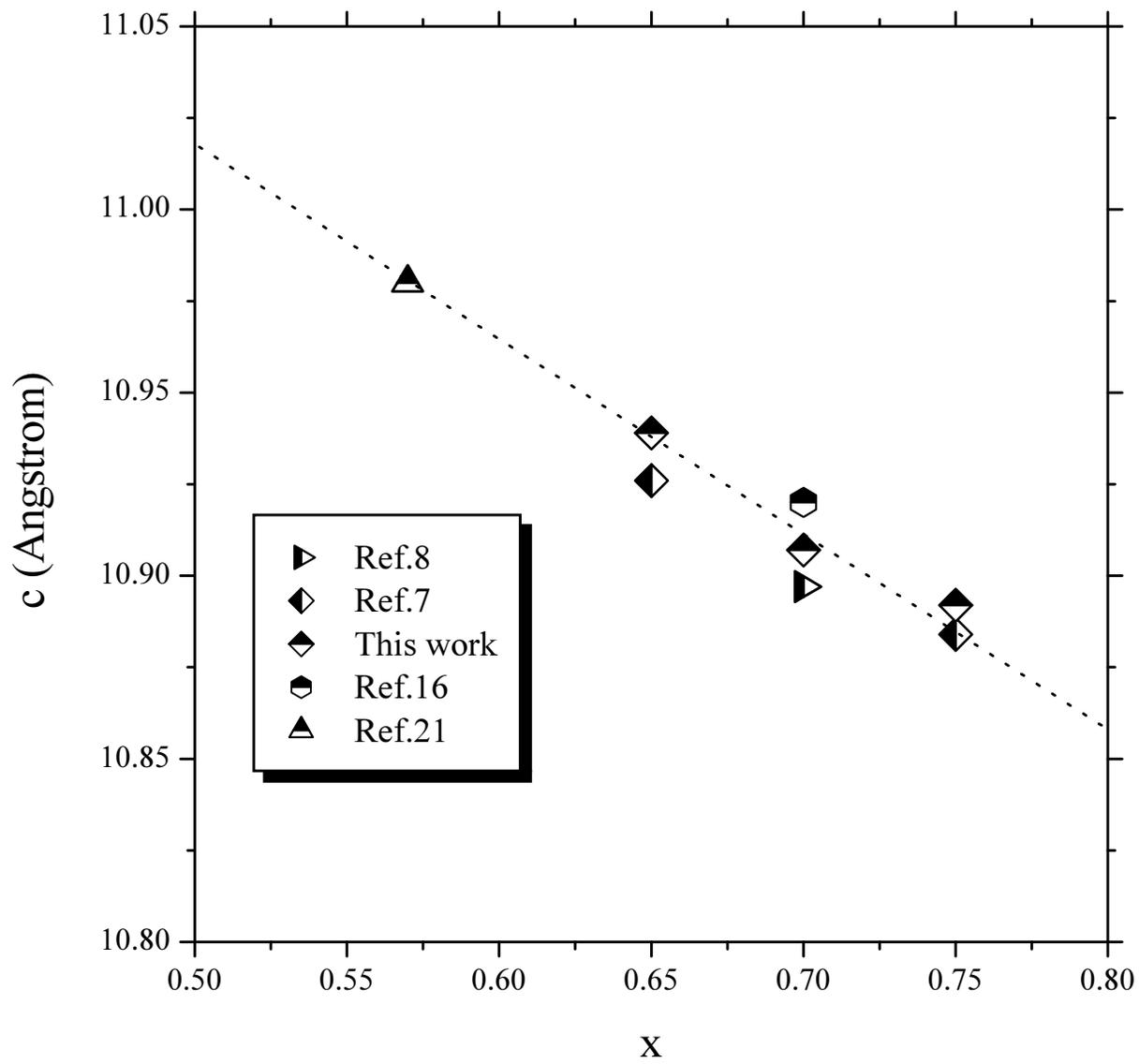
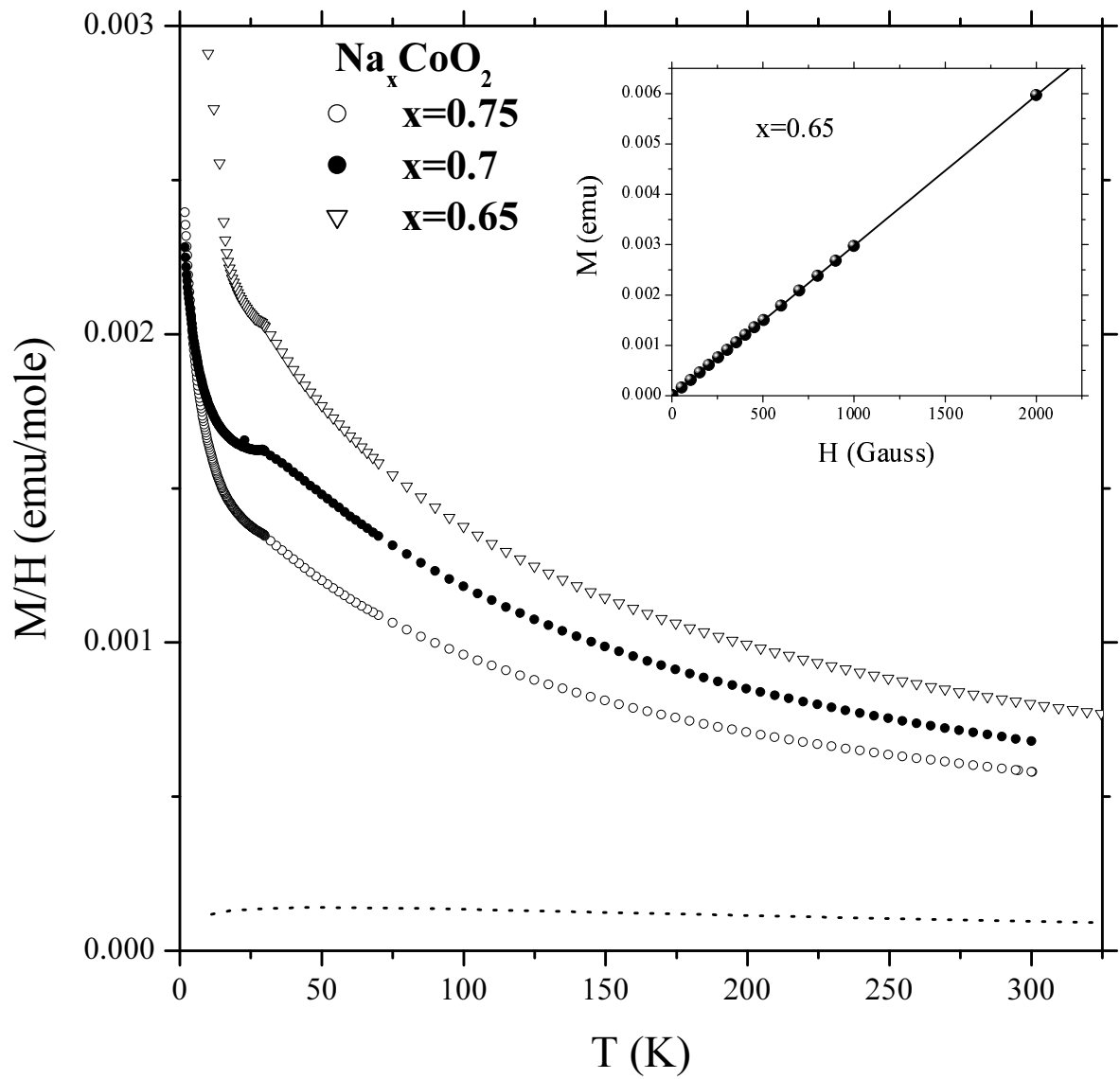
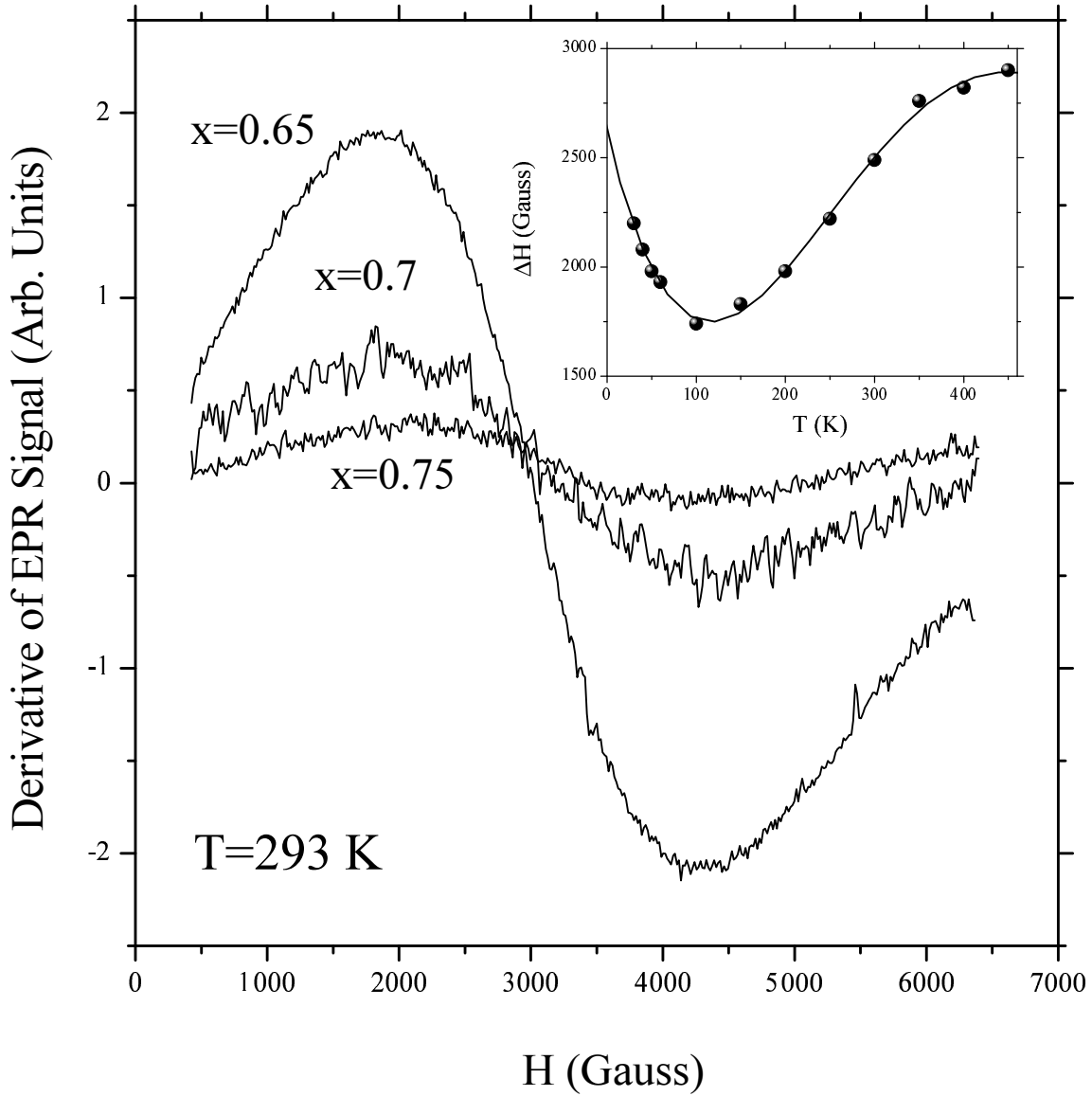
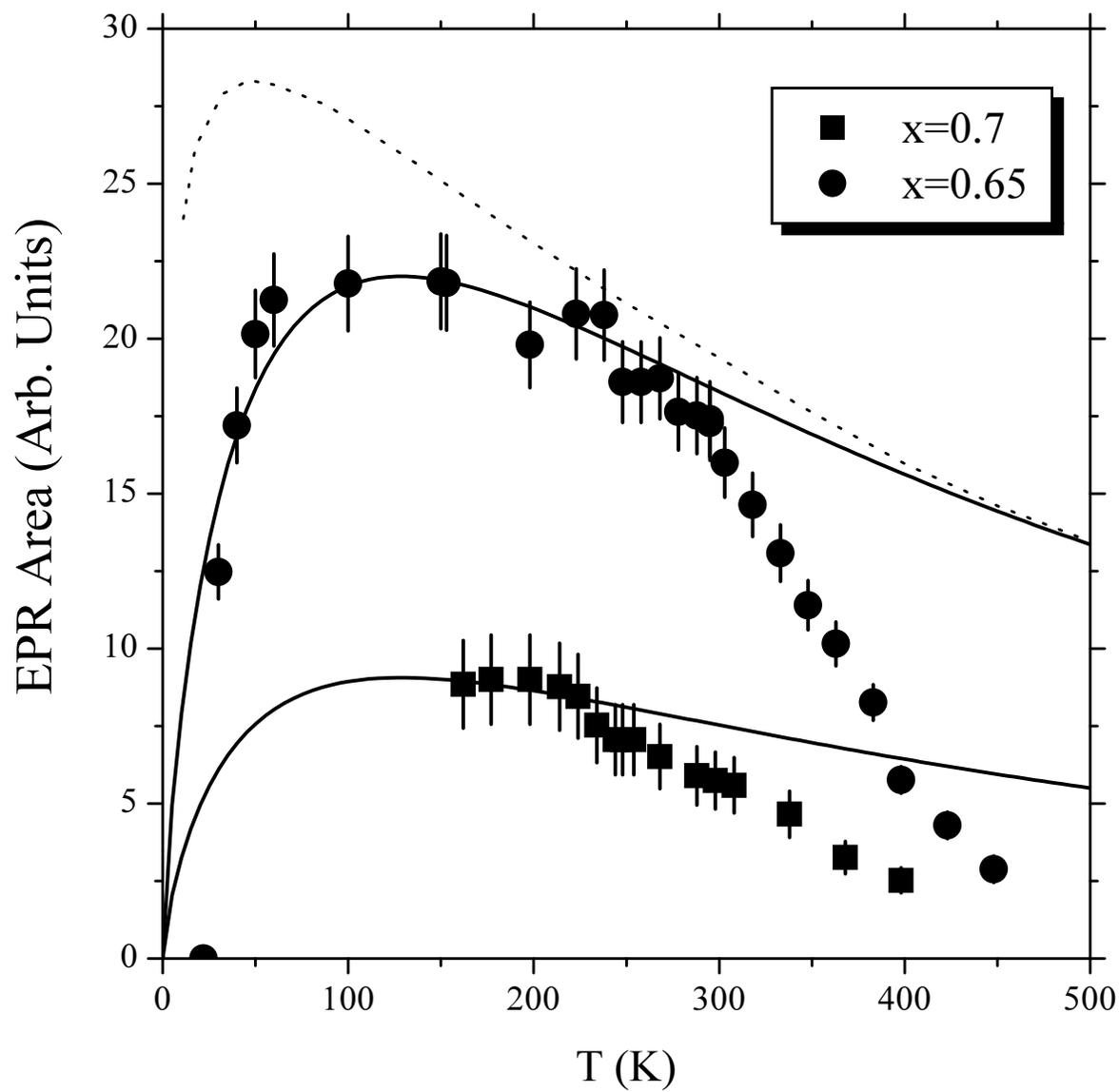
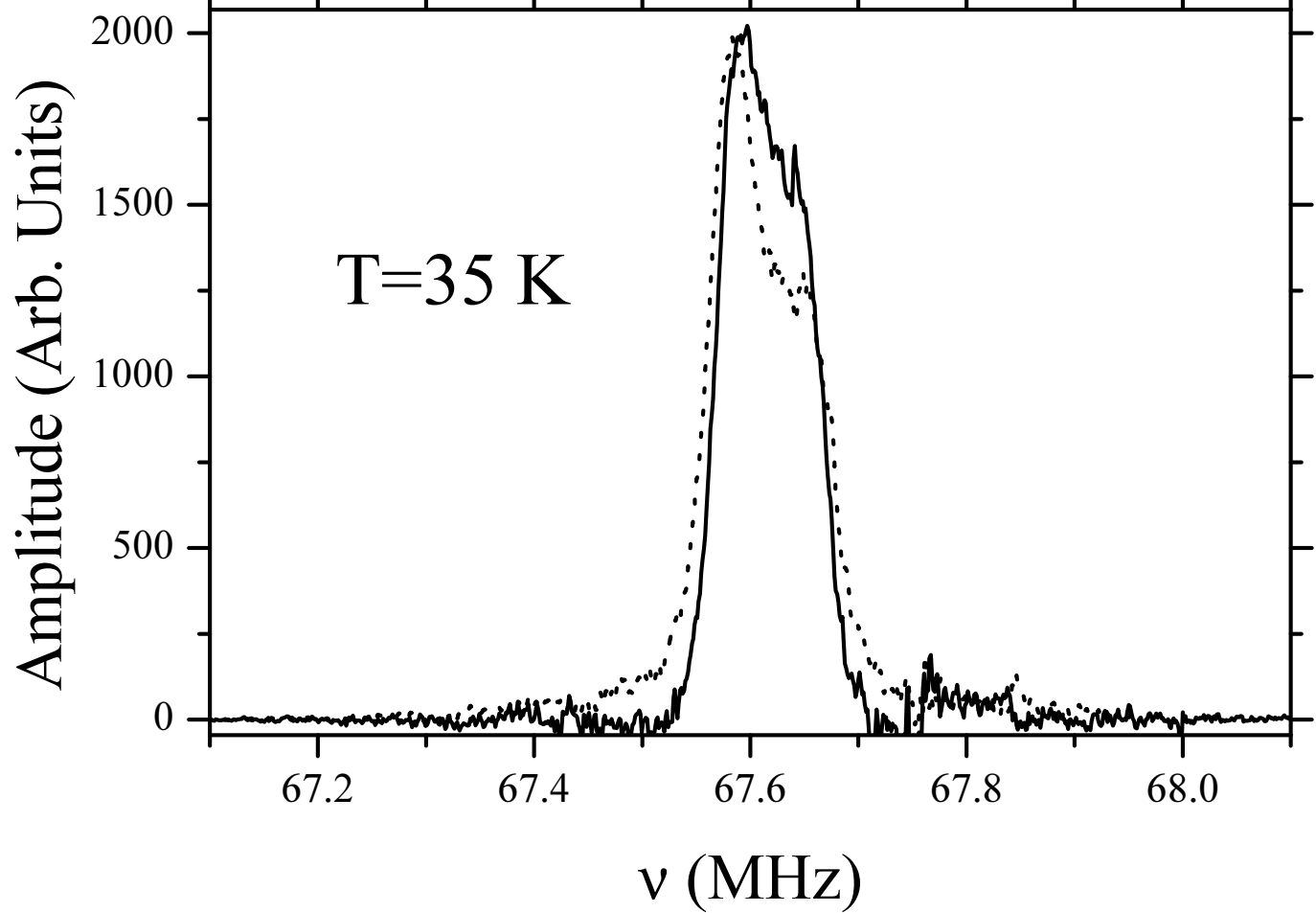
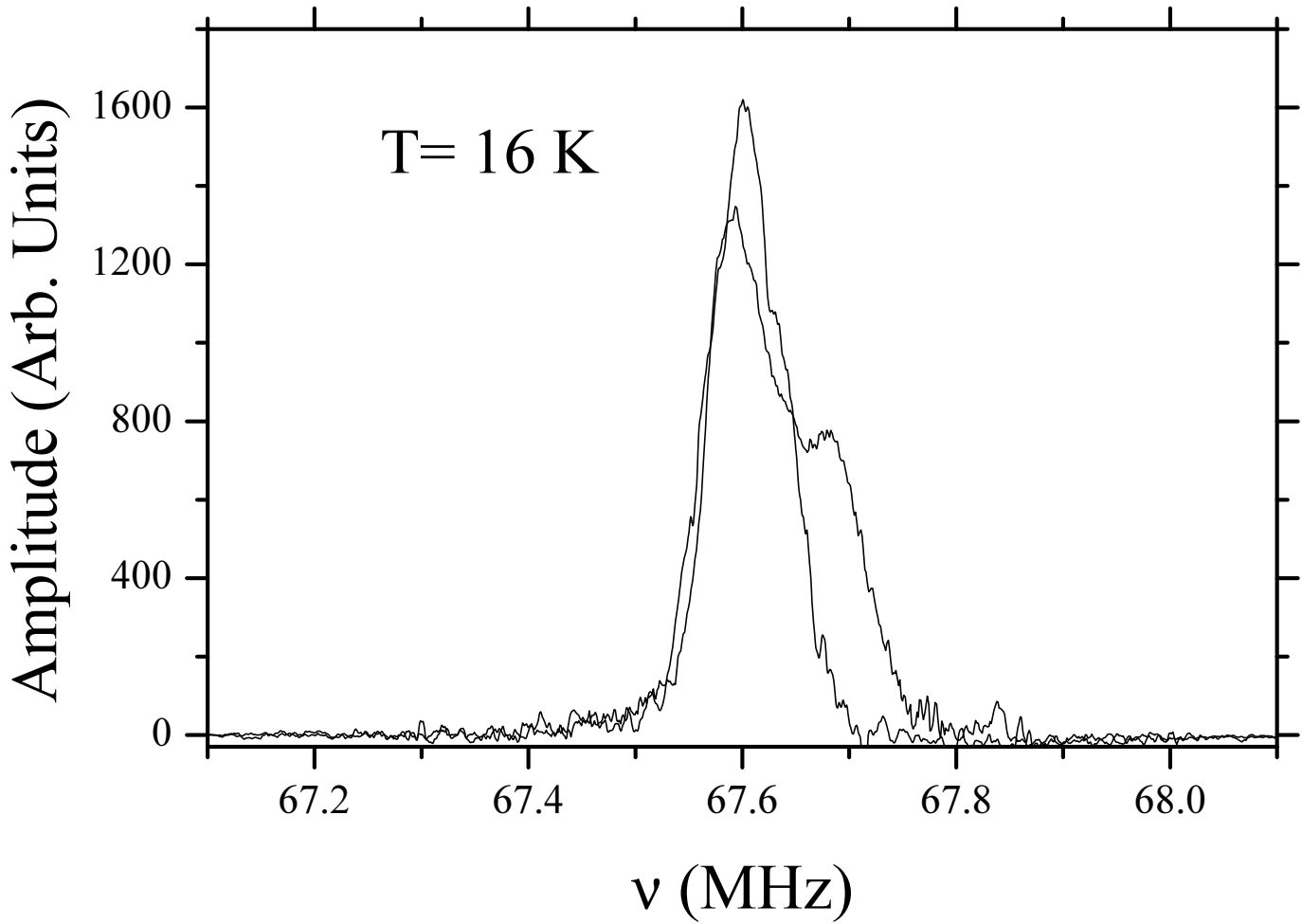
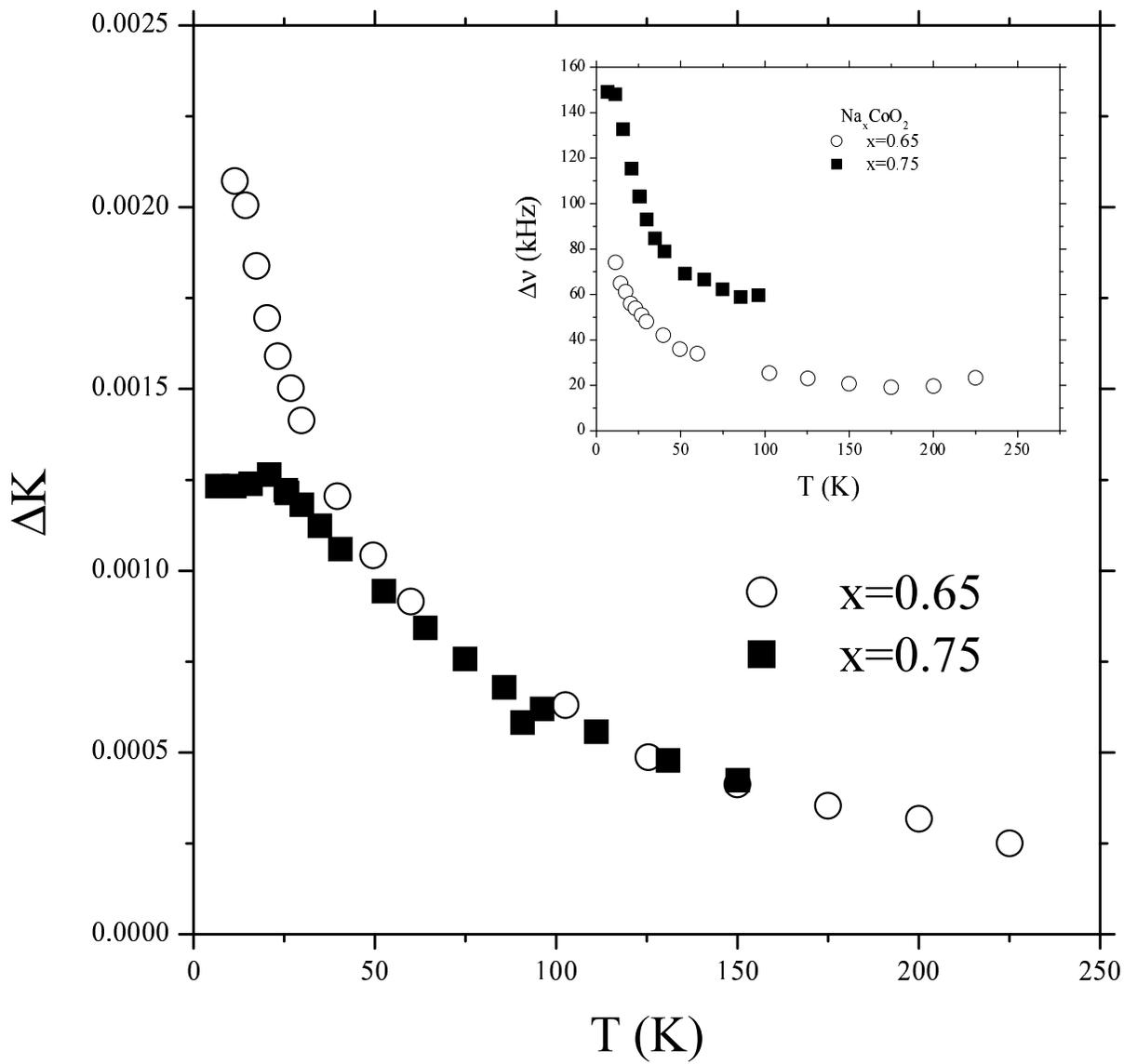
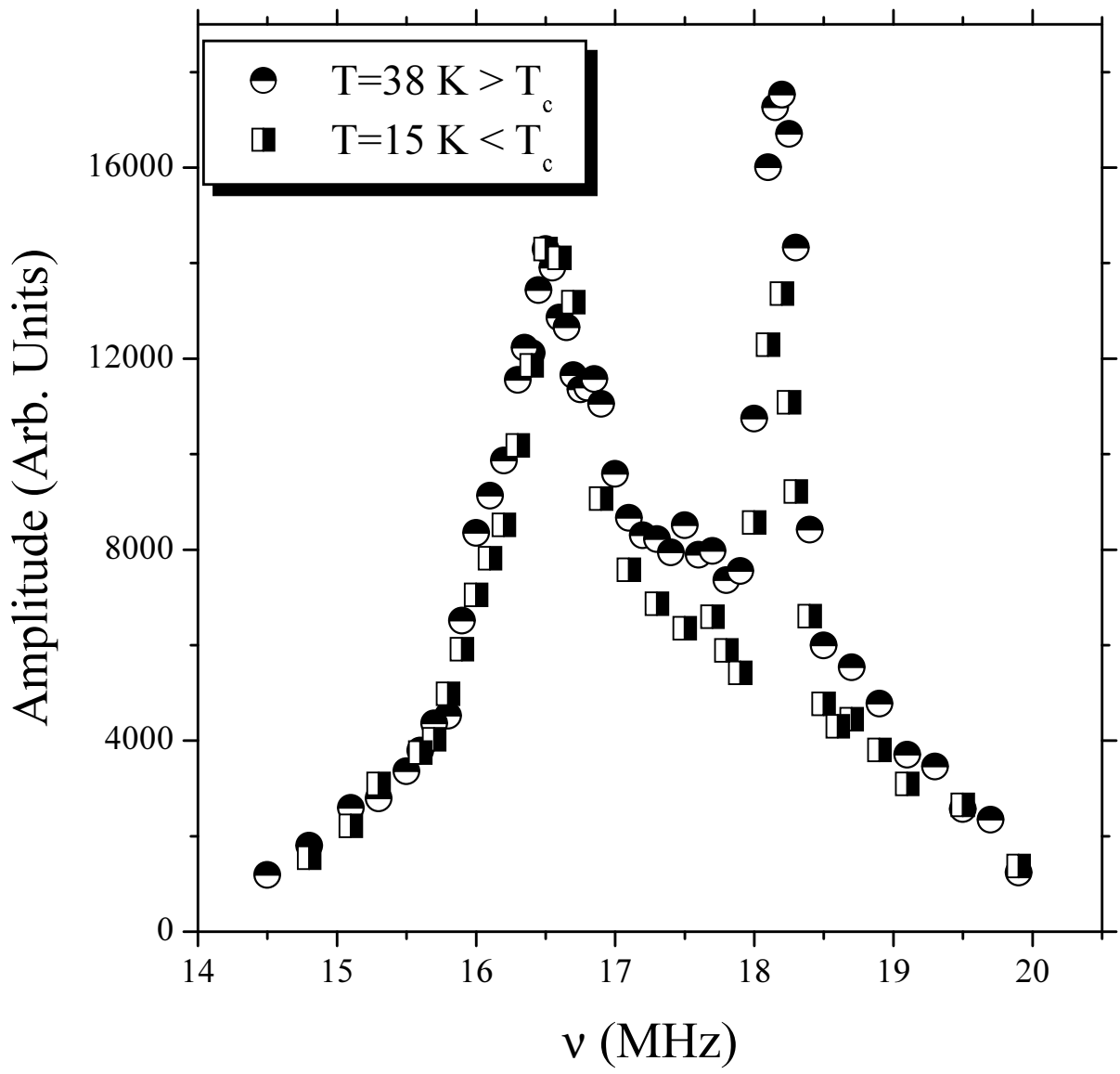
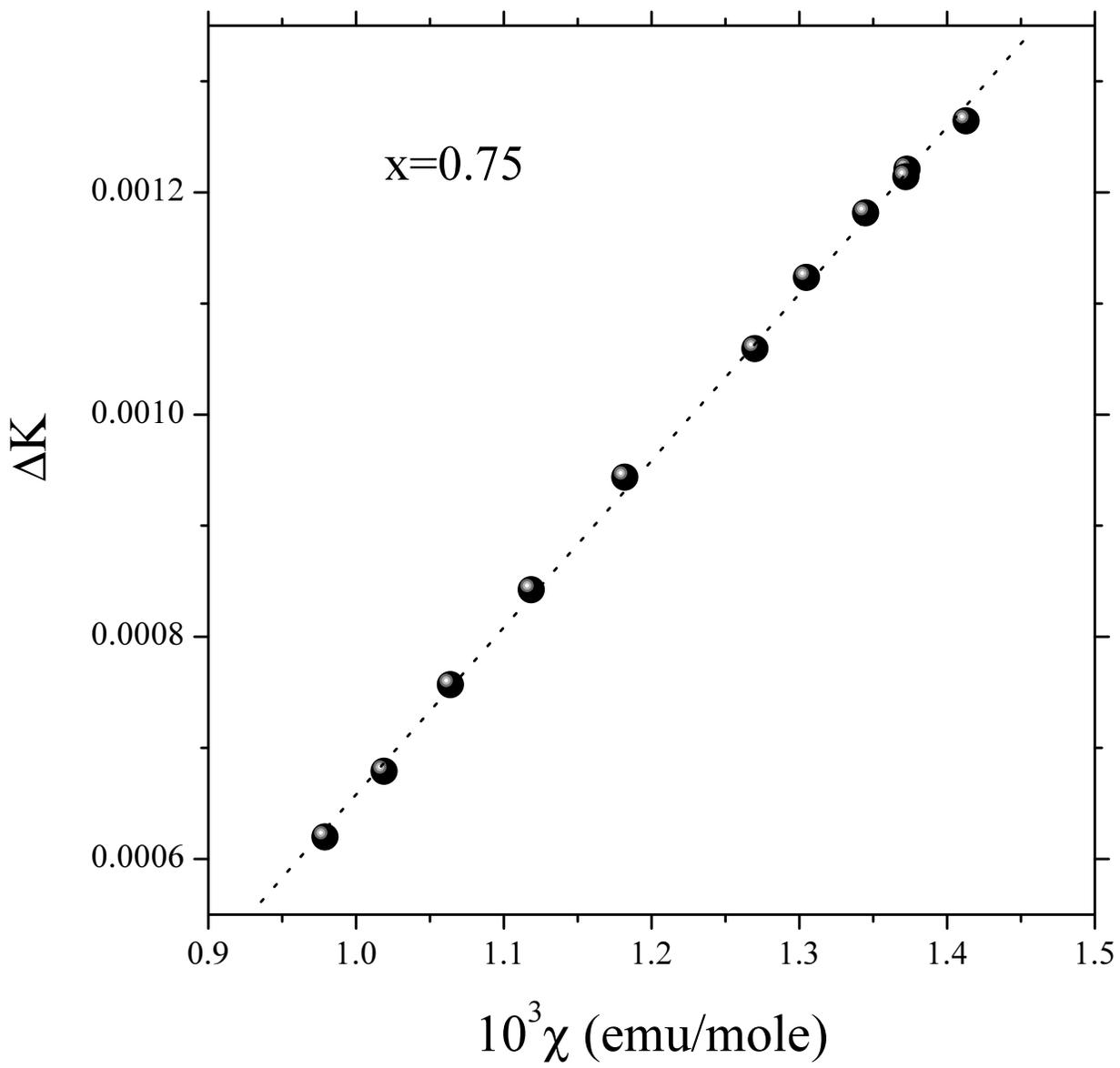
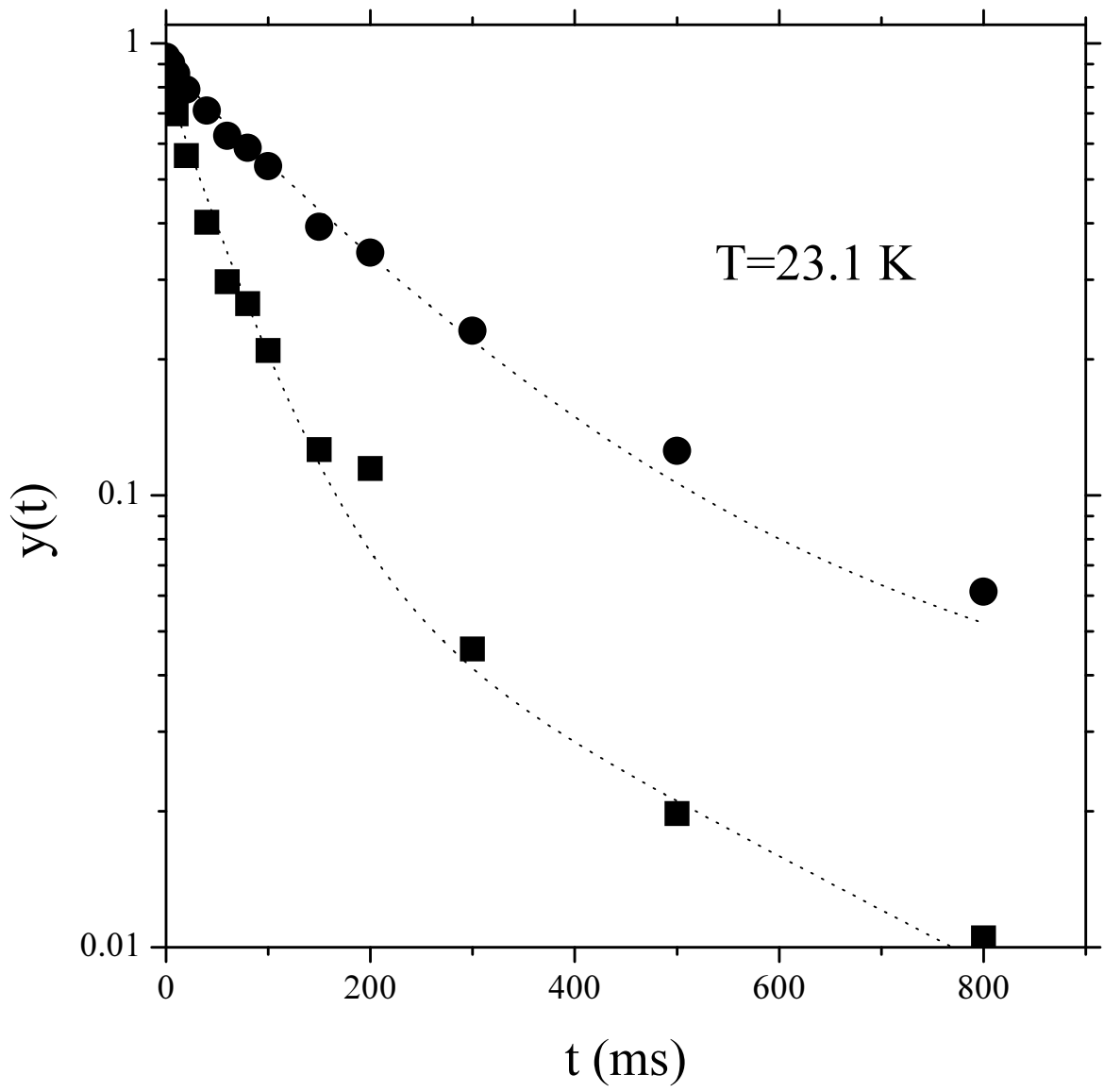
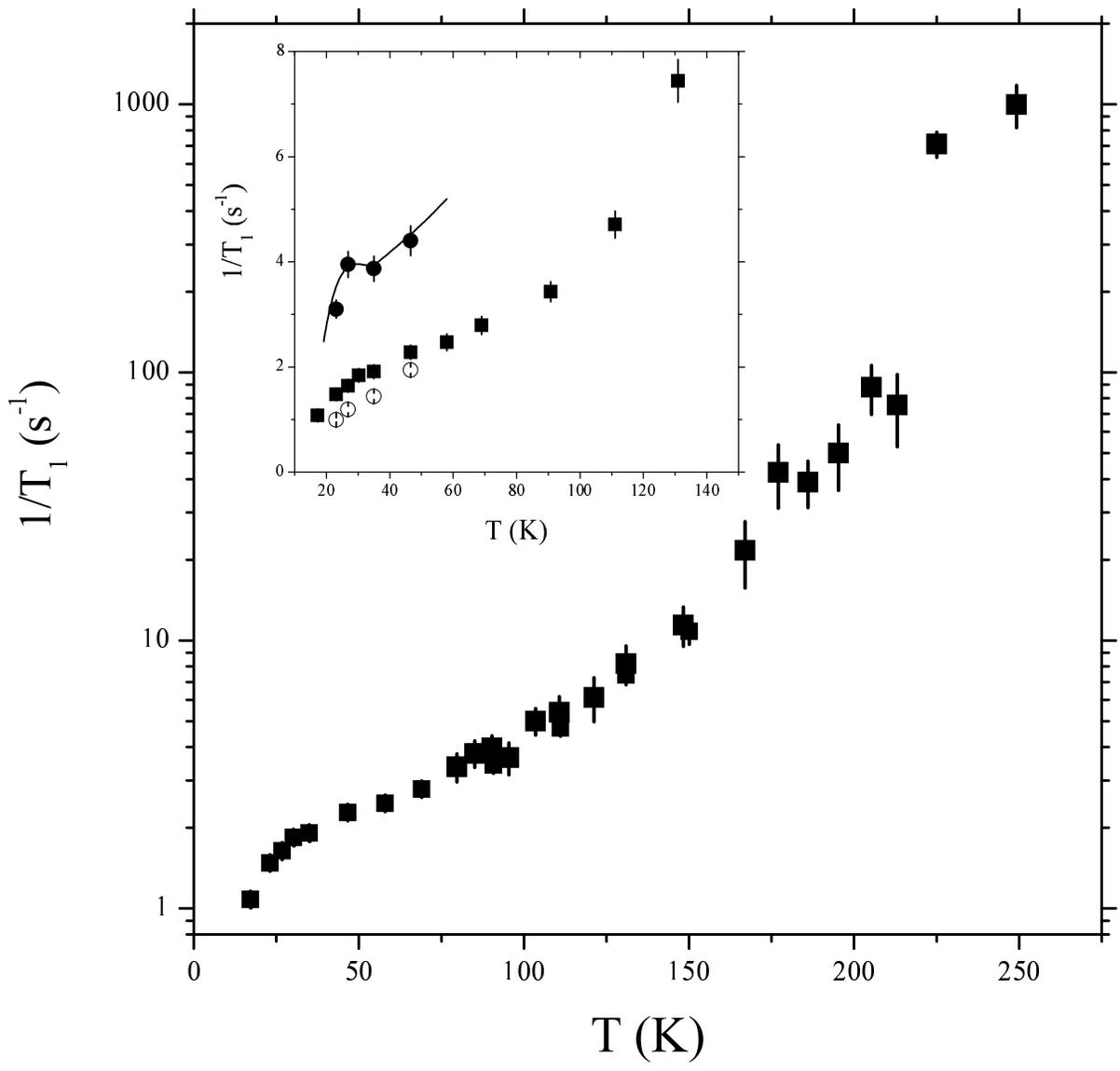
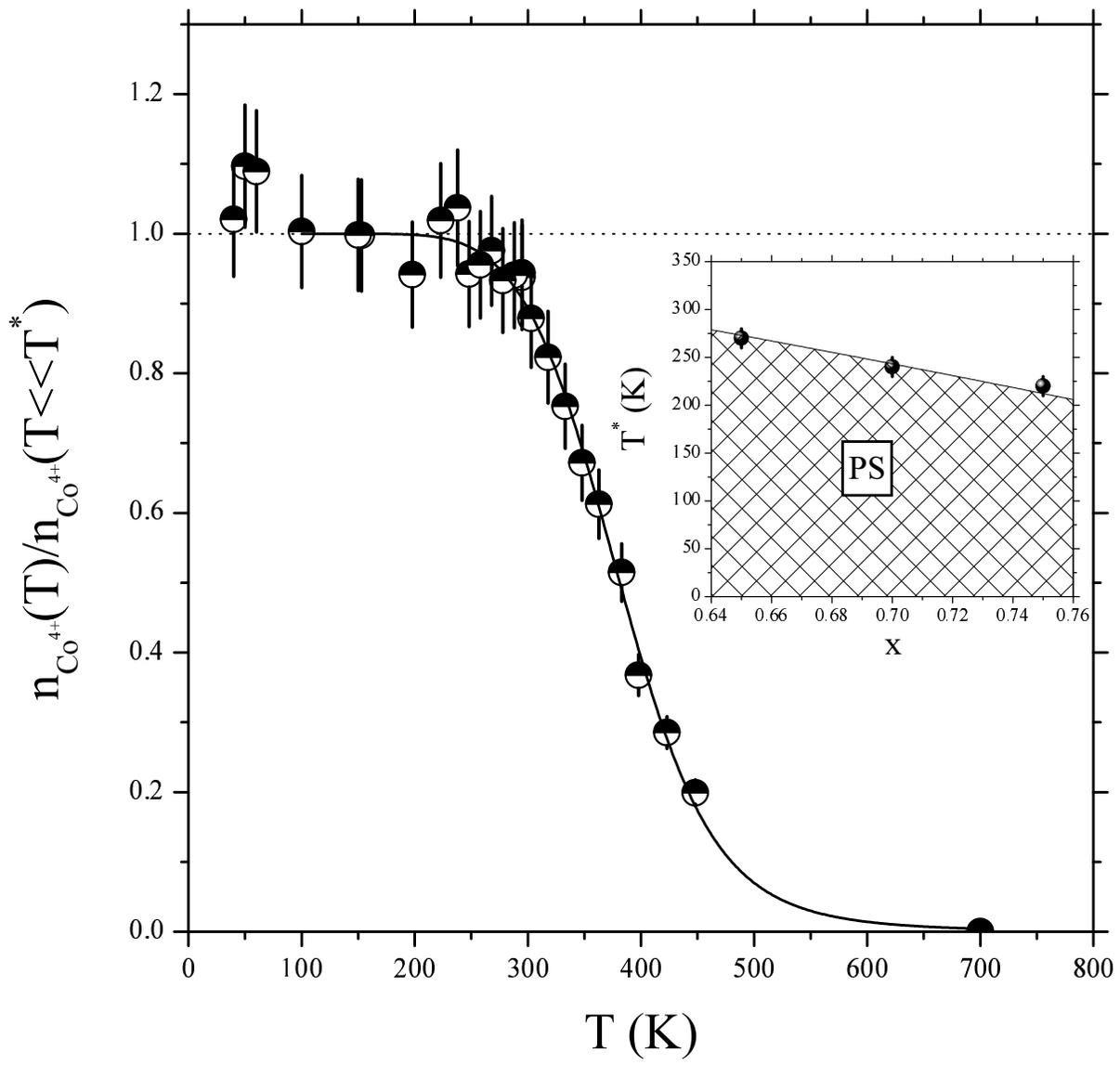
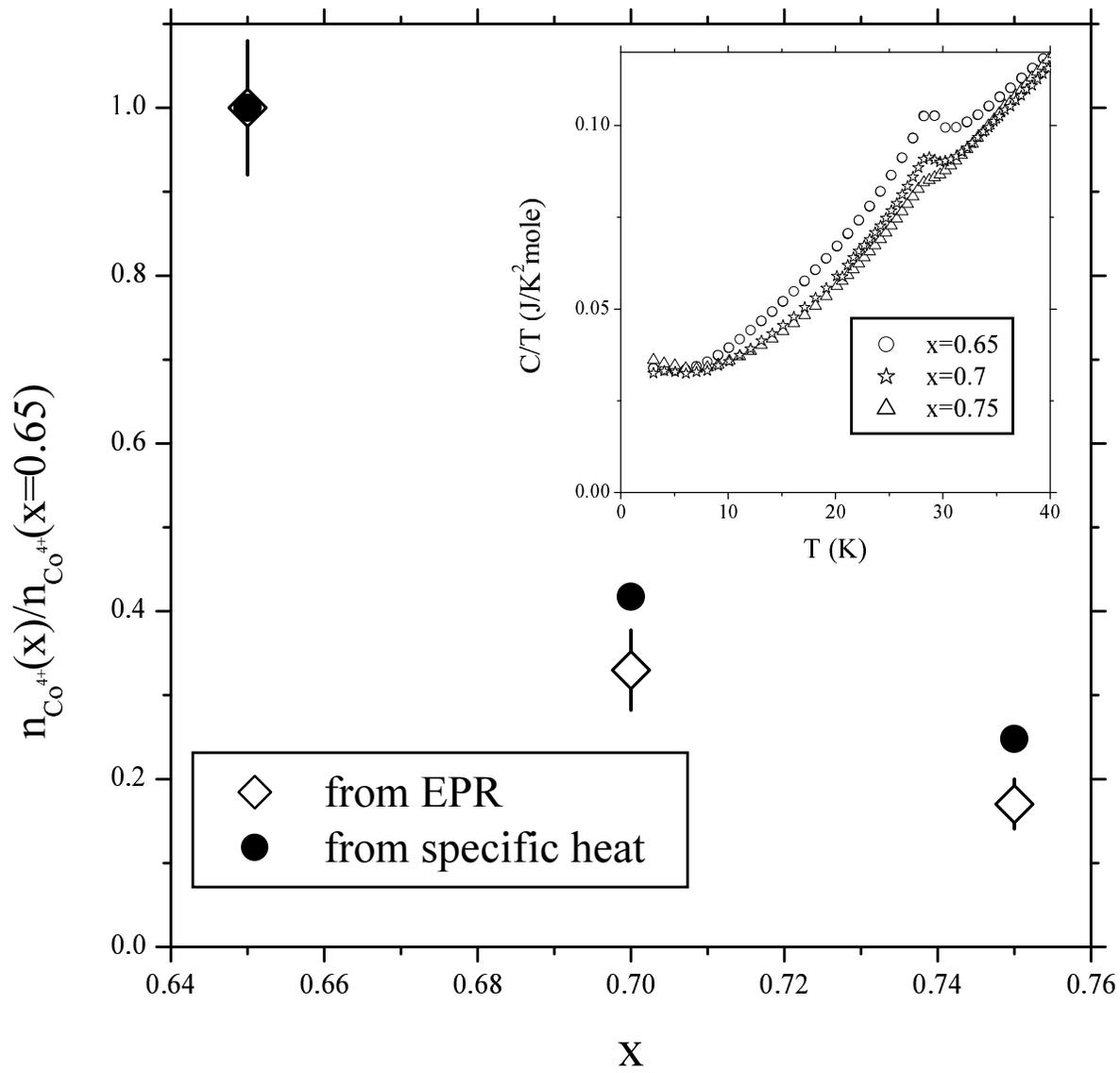
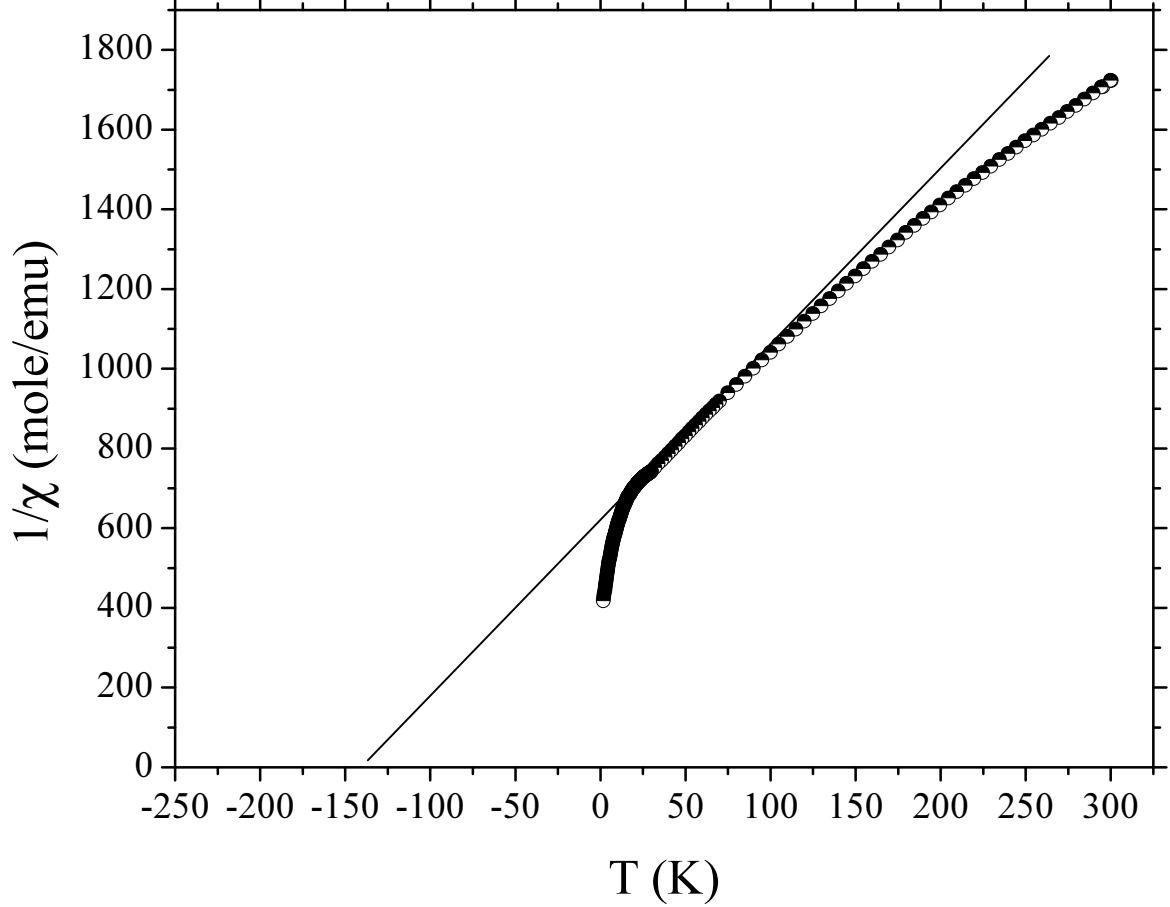
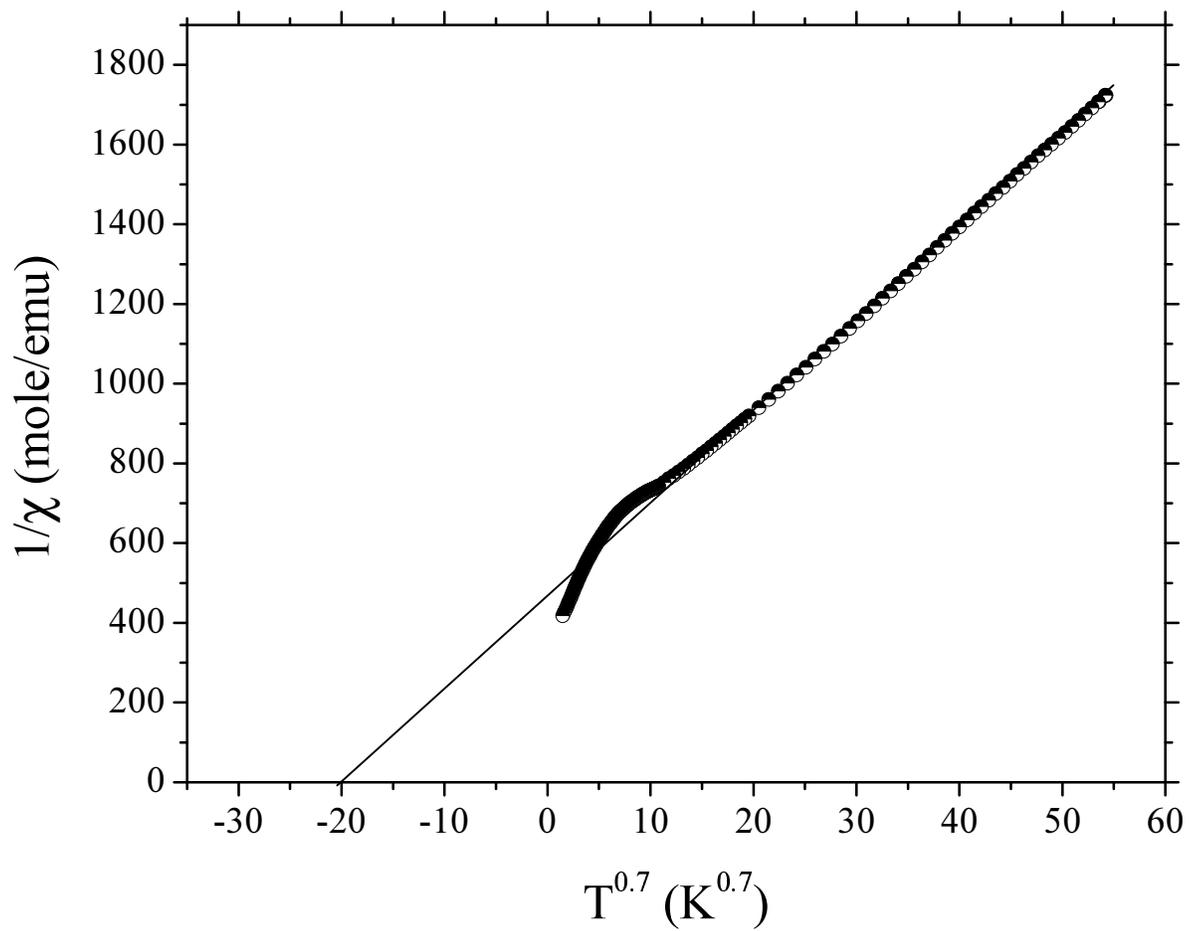
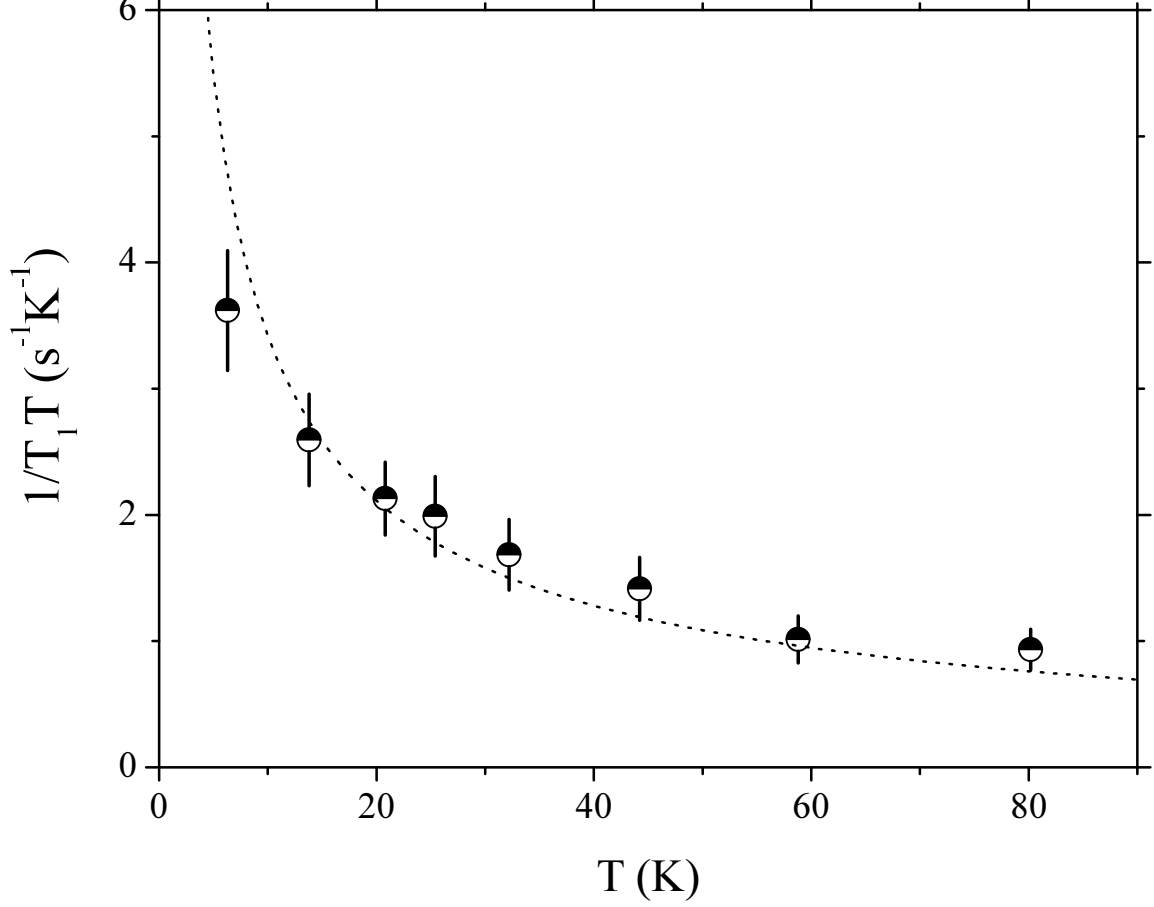
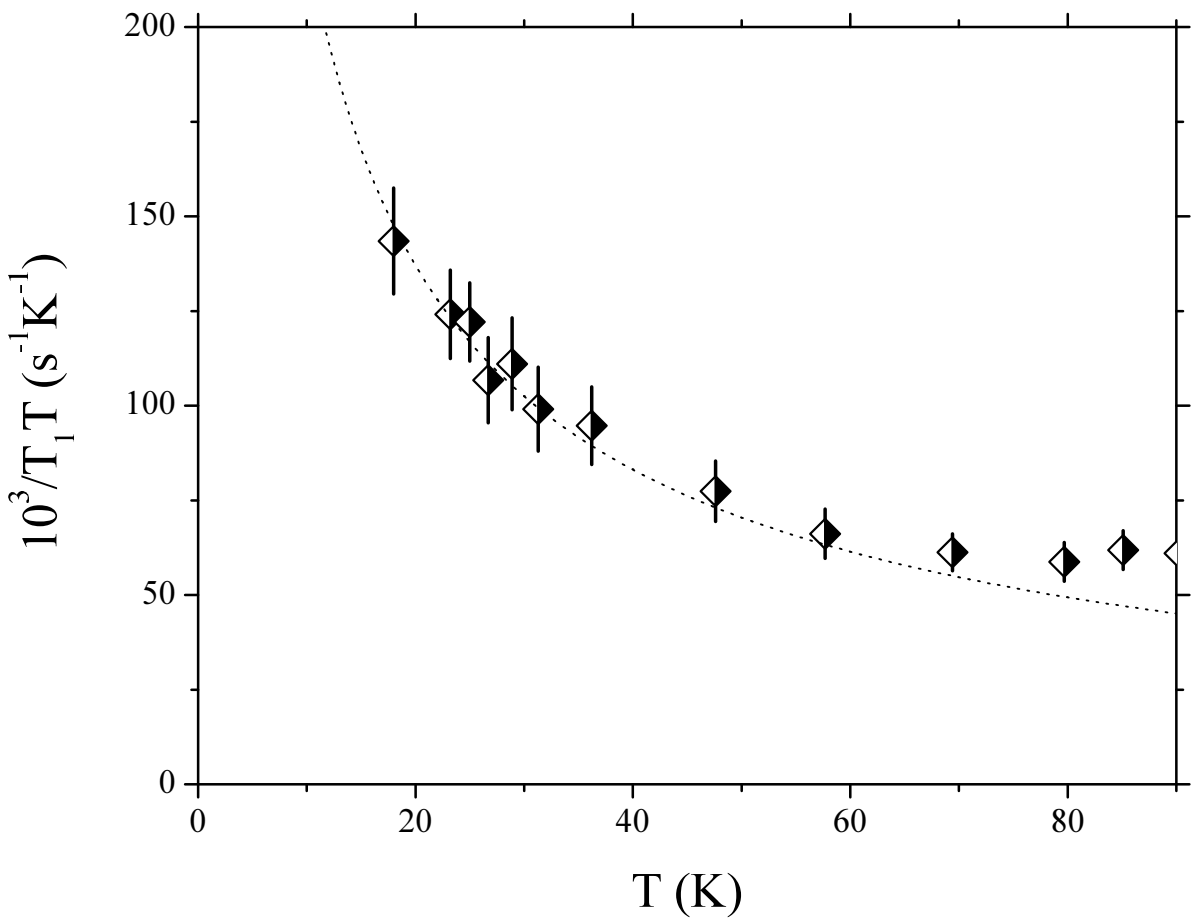
\end{document}